\documentclass[final,1p,times]{elsarticle} 
\usepackage{graphicx}
\usepackage{amssymb} 
\usepackage{amsthm} 
\newcommand{\sqrtsNN}{\sqrt{s_{\rm \scriptscriptstyle NN}}}

\newcommand{\mum}{\mathrm{\mu m}}

\newcommand{\RAA}{R_{\rm AA}}
\newcommand{\TAA}{T_{\rm AA}}

\newcommand{\Dstophipi}{{\rm D_s^{+}\to \phi\pi^+}}

\newcommand{\phitoKK}{{\rm \phi\to  K^-K^+}}

\newcommand{\Dzero}{{\rm D^0}}
\newcommand{\Ds}{{\rm D^{+}_{s}}}
\newcommand{\Dstar}{{\rm D^{*+}}}
\newcommand{\Dplus}{{\rm D^+}}

\newcommand{\dNAA}{{\rm d}N_{\rm AA}}
\newcommand{\sigmapp}{\sigma_{\rm pp}}
\newcommand{\dsigmapp}{{\rm d}\sigma_{\rm pp}}
\newcommand{\dpt}{{\rm d}p_{\rm T}}
\newcommand{\RAAFormula}{\frac{\dNAA^{h}/\dpt}{\langle \TAA \rangle \times \dsigmapp^{h}/\dpt}} %\RAA^{h}(\pt) =\frac{\dNAA^{h}/\dpt}{\langle \TAA \rangle \times \dsigmapp^{h}/\dpt}}

\journal{Nuclear Physics A} 

\begin{document}

\begin{frontmatter} 

% Your Title - please insert
\title{$\rm D_s^+$ production at central rapidity in Pb--Pb collisions \\ at $\sqrt{\rm s_{NN}}$ = 2.76 TeV
with the ALICE detector}

\author{Gian Michele Innocenti for the ALICE\fnref{col1} Collaboration}
\fntext[col1] {A list of members of the ALICE Collaboration and acknowledgements can be found at the end of this issue.}
\address{Universit\`a degli Studi di Torino and INFN,  Torino, Italy}

\begin{abstract} 

We present the measurement of the $\Ds$ production in pp collisions 
at $\sqrt{\rm s}$ = 7 TeV and 
in Pb--Pb collisions at $\sqrtsNN$ =2.76 TeV
performed with the ALICE detector at central rapidity through the reconstruction 
of the hadronic decay channel $\rm D^{+}_{s} \rightarrow \phi \pi^{+} \rightarrow K^{+}K^{-}\pi^{+}$.
The preliminary results of the $\Ds$ nuclear modification factor will be presented.
 
\end{abstract} 

\end{frontmatter} % do not change

%% linenumbers are useful for reviewing process
%\linenumbers

\section{Introduction}
The measurement of heavy-flavour production provides insights on the properties 
of the high-density QCD medium created in ultra-relativistic heavy-ion collisions.
The comparison of charm production in pp and in Pb--Pb collisions allows one 
to study the mechanism of in-medium partonic energy loss of heavy 
quarks which are expected to lose less energy with respect 
to light partons due to the presence of the so called "dead cone" effect~\cite{energyloss}.
In addition, since strange quarks are abundant in the medium~\cite{strange}, the
relative yield of $\Ds$-mesons with respect to non-strange charm mesons 
($\Dzero$, $\Dplus$ and $\Dstar$) is predicted to be largely 
enhanced if in-medium hadronization is the dominant mechanism for charm hadron formation 
in the low momentum region~\cite{rapp,rafelsky}.
A sensitive observable in order to address these open questions is the nuclear modification 
factor, which is defined for a given particle species $h$ as $\RAA^{h}(p_{\rm T})$= $\RAAFormula$,
where $N^{h}_{\rm AA}$ is the yield measured in heavy-ion collisions, 
$\langle \TAA \rangle$ is the average nuclear overlap
function calculated using the Glauber model and $\sigmapp^{h}$ is the production 
cross section in pp collisions at the same energy. 
In these proceedings we present the first measurement of the nuclear modification 
factor of the $\Ds$-meson in Pb--Pb collisions at 2.76~TeV
performed in the central rapidity region of the ALICE experiment. 
A detailed description of the pp results used as a reference for this analysis
can be found in~\cite{Ds}.
\section{Detector layout and data sample}

Open charm measurements are performed in ALICE using the tracking detectors and
particle identification systems of the central barrel, which covers the pseudo-rapidity 
region -0.9 $< \eta <$ 0.9 and is embedded in a magnetic field of B = 0.5 T.
The central barrel detectors allow to track charged meson down 
to low transverse momenta ($\approx$ 100 MeV/c) and provides charged hadron and electron 
identification as well as an accurate measurement of the positions of the primary and secondary 
(decay) vertices~\cite{alice}.
A short description of the detectors utilized in these analyses will be given in this section. 
The Time Projection Chamber (TPC) is the main tracking detector which provides track reconstruction 
and particle identification via the measurement of the specific energy deposit 
${\rm d}E/{\rm d}x$. The Inner Tracking System (ITS) is the central barrel 
detector closest to the beam axis and it is composed of six cylindrical layers of silicon detectors.
The two innermost layers (at radii of 3.9 and 7.6 cm) are made of pixel detectors (SPD), 
the two intermediate layers (radii $\approx$ 15 and 24 cm) are equipped with drift detectors, 
while strip detectors are used for the two outermost layers (radii $\approx$ 39 and 44 cm).
The ITS is crucial to reconstruct secondary vertices 
originating from open charm decays because it allows for the measurement of the 
track impact parameter in the transverse plane with a resolution better than 50 $\mu$m for tracks with $p_{\rm T}$ $>$ 1.3 GeV/$c$~\cite{paperoD}. 
The Time-of-Flight (TOF) detector is used for pion, kaon and proton identification 
on the basis of their time of flight. The TOF measurement provides kaon/pion separation up to a 
momentum of about 1.5 GeV/$c$. All the three detectors have full azimuthal coverage. 
The data sample used for this analysis has been collected during the 2011 LHC run (November-December 2011)
with Pb--Pb collisions at $\sqrt{s_{\rm NN}}$ = 2.76 TeV and it consists of 16 million events  
selected in the centrality range 0 -- 7.5$\%$ according to centrality-based trigger requirements 
on the signal from the SPD and from the VZERO detectors. 
The latter is made of two scintillator hodoscopes positioned in the forward and backward regions 
of the experiment~\cite{alice}. 
Only events with a vertex found within $\pm$ 10 cm from the centre of the detector along 
the beam line were used for  the analysis.

\section{$\Ds$ meson reconstruction in ALICE}
\begin{figure}[htbp]
\begin{center}
\includegraphics[width=1.\textwidth]{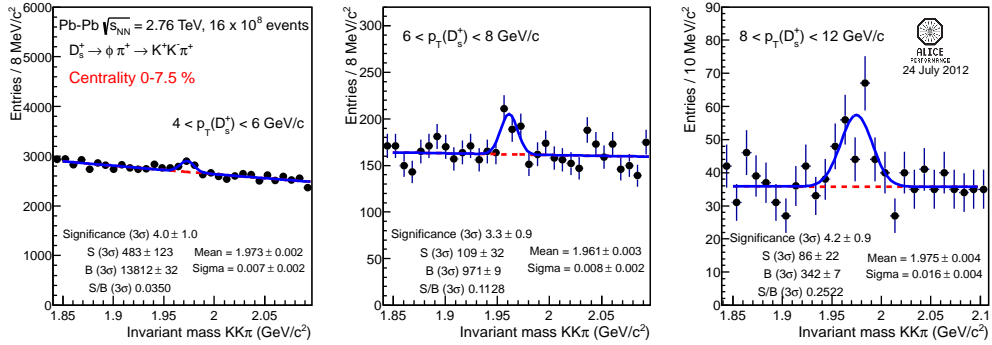}
\caption{Invariant mass distributions for $\Ds$ candidates and charge
conjugates in the three considered $p_{\rm T}$ intervals in the range 4 $<$ $p_{\rm T}$ $<$ 12 GeV/$c$
obtained from the analysis of 16 million Pb--Pb events in the centrality range 0 -- 7.5$\%$.}
\label{fig:invmass}
\end{center}
\end{figure}

$\Ds$ mesons and their antiparticles were reconstructed in the 
decay chain $\Dstophipi$ (and its charge conjugate) followed by $\phitoKK$. 
The branching ratio (BR) of this chain is $2.28 \pm 0.12\%$~\cite{PDG}. $\Ds$ mesons have a mean proper decay length 
$c \tau=150 \pm 2~\mum$~\cite{PDG} which allows to resolve their decay vertex 
from the interaction (primary) vertex. Thus, the analysis strategy for the extraction 
of the signal out of the large combinatorial background can be based on the reconstruction 
and selection of secondary vertex topologies with significant separation from the primary vertex. 
The $\Ds$ meson candidates are built starting from track combinations
with proper charges and selected according to topological cuts. Single tracks are previously selected
with respect to to their pseudorapidity ($|\eta|$ $<$ 0.8), momentum and to quality cuts.
%($p_{\rm T}$ $>$ 0.4 GeV/$c$ for the pp analysis and $p_{\rm T}$ $>$ 0.7 GeV/$c$ in the Pb--Pb case)
Candidates are then filtered by applying kinematical and topological cuts 
together with particle identification criteria in order to improve the statistical 
significance while keeping the selection efficiency as high as possible. 
The candidate triplets were selected according to the 
sum of the distances of the decay tracks to the reconstructed decay vertex, 
the decay length and the cosine of the pointing angle, 
that is the angle between the reconstructed $\Ds$ meson momentum and the line connecting the 
primary and the secondary vertex. In addition, $\Ds$ candidates were selected by requiring that one
of the two pairs of opposite-charge tracks has an invariant mass compatible 
with the PDG value for the $\phi$ meson mass~\cite{PDG}. 
The same topological and PID strategy has been adopted in the pp and Pb--Pb analyses with 
tighter topological selection cuts used in nucleus-nucleus collisions to cope with the higher combinatorial background.
In Fig.~\ref{fig:invmass} the invariant mass distributions of $\Ds$ candidates in the centrality interval 
0 -- 7.5$\%$ in three $p_{\rm T}$ bins from 4 to 12 GeV/$c$ are presented. 
To evaluate the signal yield, a fit to the distributions 
with a Gaussian function for the signal and an exponential shape for the background
was used.

The measured raw yields were corrected for acceptance and selection efficiency 
of prompt $\Ds$ mesons using Monte Carlo simulations based on PYTHIA (Perugia-0 tuning)~\cite{pythia} 
and HIJING~\cite{hijing}. The correction factor that accounts for the feed-down from 
B meson decays was evaluated using the Monte Carlo efficiency for feed-down 
$\Ds$ mesons and the FONLL calculations, which describe well bottom production 
at Tevatron~\cite{teva} and at the LHC~\cite{lhc1,lhc2}. 
In the Pb--Pb analysis, the contribution estimated with FONLL was 
multiplied by the average number of binary collisions estimated via the 
Glauber Model $\langle N_{\rm coll}\ \rangle$ and by the hypothesis on the energy loss
of $\Ds$ from B ($\RAA^{\rm B}$) (more details in ~\cite{piombo}).
The systematic uncertainties on the B feed-down subtraction was 
estimated as the full spread of the results obtained by varying the FONLL 
parameters and  the values of $\RAA^{\rm B}$ (the latter only in the Pb--Pb case).
In particular, the hypothesis on the $\RAA^{\rm B}$ has been varied in the range
1/3 $<$ $\RAA^{\rm B}$/$\RAA^{\rm D}$ $<$ 3. The estimated fraction of $\Ds$
from B meson decays is $\approx$ 0.8 in the three $p_{\rm T}$ intervals considered.
In order to obtain a reference for the nuclear modification factor, the pp 
cross section at 7 TeV has been scaled to $\sqrt{s}$ = 2.76 TeV using the ratio
of the FONLL cross sections for the $\Ds$ meson production at the two energies.
To compute the systematic uncertainties on the scaling, 
the parameters of the FONLL calculation (factorization and renormalization scales,
charm quark mass) were varied. The resulting contribution to the systematic uncertainty
is about 15$\%$. \\ The validity of the energy scaling was tested for the non-strange 
$\Ds$ meson through the comparison with the measured $p_{\rm T}$-differential cross section
in pp collisions at 2.76 TeV~\cite{low}.

\section{Results}
\begin{figure}[htbp]
\begin{center}
\includegraphics[width=0.40\textwidth]{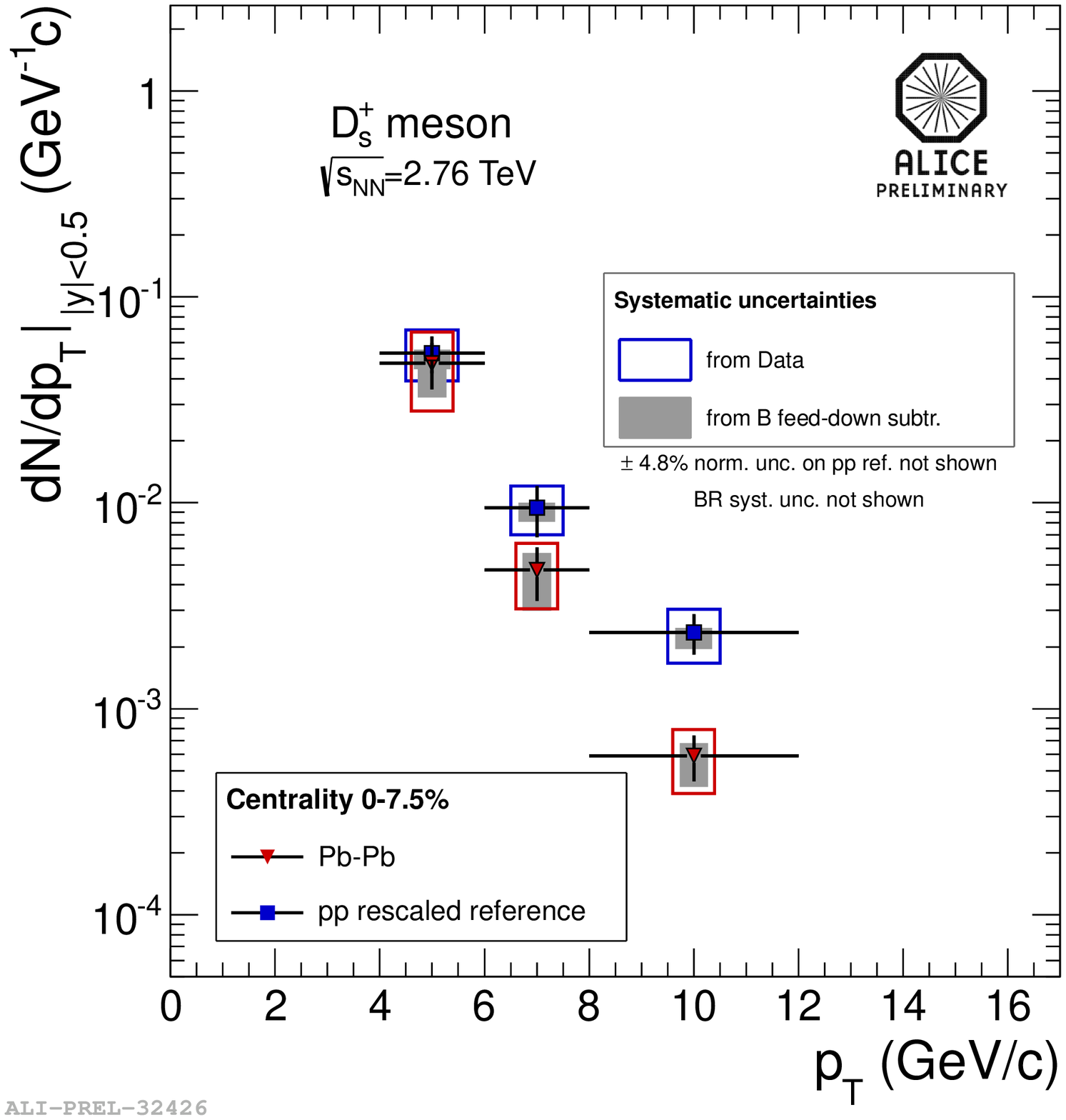}
\hspace{0.7cm}
\includegraphics[width=0.40\textwidth]{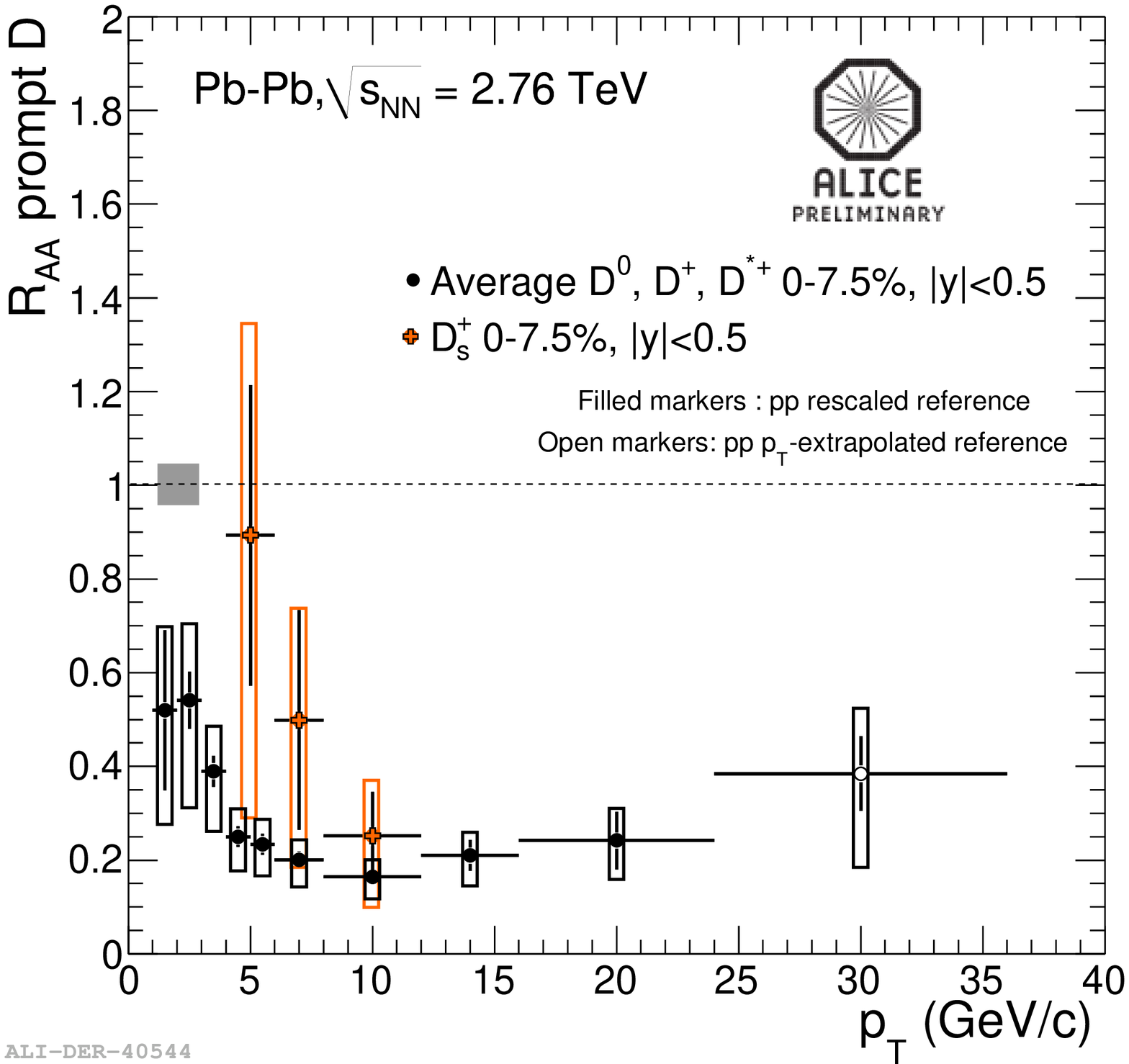}
\caption{(Left) $\Ds$ yield for central Pb--Pb collisions (0 -- 7.5$\%$) at $\sqrtsNN$=2.76 TeV compared
to the reference spectra rescaled at the same energy. (Right) $\Ds$ nuclear modification factor 
measured in central (0 -- 7.5$\%$) Pb--Pb collisions compared to the average non-strange D-meson
$\RAA$ also measured in ALICE~\cite{zaida}.}
\label{fig:resultsPbPb}
\end{center}
\end{figure}

In Fig. \ref{fig:resultsPbPb} (Left) the $\Ds$ yield measured in Pb--Pb collisions
at $\sqrt{s_{\rm NN}}$ =2.76 TeV in the centrality range 0 -- 7.5$\%$ and
the rescaled pp reference at the same energy are presented. In the right panel of Fig. \ref{fig:resultsPbPb}
the $\Ds$ $\RAA$ in the same centrality class is shown and compared to the average non-strange D-meson
one ($\Dzero$, $\Dplus$ and $\Dstar$)~\cite{zaida}. The vertical bars represent the statistical uncertainties while the 
total systematic uncertainties are shown as boxes. The total systematics 
include the uncertainties on signal extraction, track reconstruction
efficiency, cut and PID selection and on the B feed-down subtraction. 
The last contribution is e.g. ${}^{+11\%}_{-45\%}$ in the transverse momentum range 4 $<$ $p_{\rm T}$ $<$ 6 GeV/$c$
and the total systematic uncertainty is ${}^{+50\%}_{-68\%}$.
The $\Ds$ nuclear modification factor in the highest $p_{\rm T}$ bin is $\approx$ 0.25 and it is compatible 
within the uncertainties with the value measured for non-strange D-mesons.
At lower transverse momenta, the $\Ds$ $\RAA$ presents an increase which is however not 
significant if compared to the average value of $\Dzero$, $\Dplus$ and $\Dstar$. Therefore, larger statistics samples 
from the future LHC runs should
allow us to draw firm conclusions on a possible enhancement of $\Ds$ production
in Pb--Pb collisions

\end{document}